\documentstyle[psfig]{l-aa}

\begin{document}

\title{NGC~6994: an open cluster which is not an open cluster\thanks{Based on observations carried out at ESO, La Silla.}}
\author{Giovanni Carraro}
\institute{Dipartimento di Astronomia, Univerit\'a di Padova, Vicolo Osservatorio 5, I-35122, Padova, 
Italy\\
e-mail: carraro@pd.astro.it}

\offprints{Giovanni Carraro, e-mail: carraro@pd.astro.it}
\date{Received: December 1999;}

\maketitle
\markboth{Giovanni Carraro}{NGC~6994}

\begin{abstract}
We report on CCD photometry in the Johnson 
$B$, $V$ and $I$ passbands for 146 stars in a $9^{\prime} \times 9^{\prime}$ 
region around the 
southern aggregate NGC~6994 (C 2056-128),
which appears in the Lyng{\aa} (1987) catalogue of open star clusters.
We argue that this object is not really
an open cluster, but simply a random enhancement
of  four bright stars above the background level. 
This stars sample includes HD~358033 and  GSC~05778--0082,
together with M~73, which is referred to as a binary or multiple star,
but actually represents the whole asterism. 
Since NGC~6994 is not the first case (see for instance Carraro \& Patat 1995),
this raises the possibility that other open clusters
may have been misclassified. We also suggest that 
NGC~6994 is unlikely  to be an open cluster remnant (OCR).
\end{abstract}


\section{Introduction}
The disk of the Milky Way harbors about 1200 open clusters.
The youngest ones trace the spiral pattern in the disk, whereas the oldest
ones are ideal templates to study the chemical and dynamical evolution
of the disk, together with the Star Formation History
(Friel 1995).\\
Open star clusters are usually studied by analyzing the Color Magnitude
Diagram (CMD, Chiosi et al 1992).\\  
Despite the importance of these objects for our understanding of the
disk properties, CMDs are at present 
available only for a small fraction of the clusters population, say 40$\%$.
However the situation is rapidly changing thanks to dedicated surveys
like for instance the PLCON (Palomar Las Campanas Observatory NOAO) Open
Clusters Survey (Phelps 1999).\\
One of the crucial problem with open clusters is that they are highly contaminated
by field stars in the disk, so that it is rather difficult to obtain
precise estimates of their fundamental parameters.\\
This way many clusters remain completely unstudied but for the identification,
which in most cases is  done on a by eye basis, by inspecting at several
different sky charts.\\
In some cases the identification is made by recognizing a peak of star
concentration in the field. This is naturally a good criterion, which
however deserves further studies to confirm the real nature of the star
enhancement.\\
This is not only a semantic question. 
A star cluster is doubtless a star enhancement in the field. \\
Nonetheless 
the real nature can be unraveled with higher degree of confidence
  by looking at the CMD, and recognizing typical
features, like a Main Sequence of H-burning stars, a Red Giant Branch,
and/or a clump of He-burning stars (Chiosi et al 1992). \\
For instance Carraro \& Patat (1995) analyzed photometry for the presumed
old cluster Ruprecht~46, and concluded that it is not a cluster, but only
a random enhancement of bright stars.\\ 
\noindent
Recently, de la Fuente Marcos (1998) has performed numerical simulations
of open clusters evolution, suggesting that many star concentrations in the sky
may be OCRs. The basic
criterion is the evidence of a star concentration higher than the field 
stars. So in principle an enhancement of stars could be just the final stage of
the evolution of an open cluster.

\noindent
\begin{figure*}
 \centerline{\psfig{file=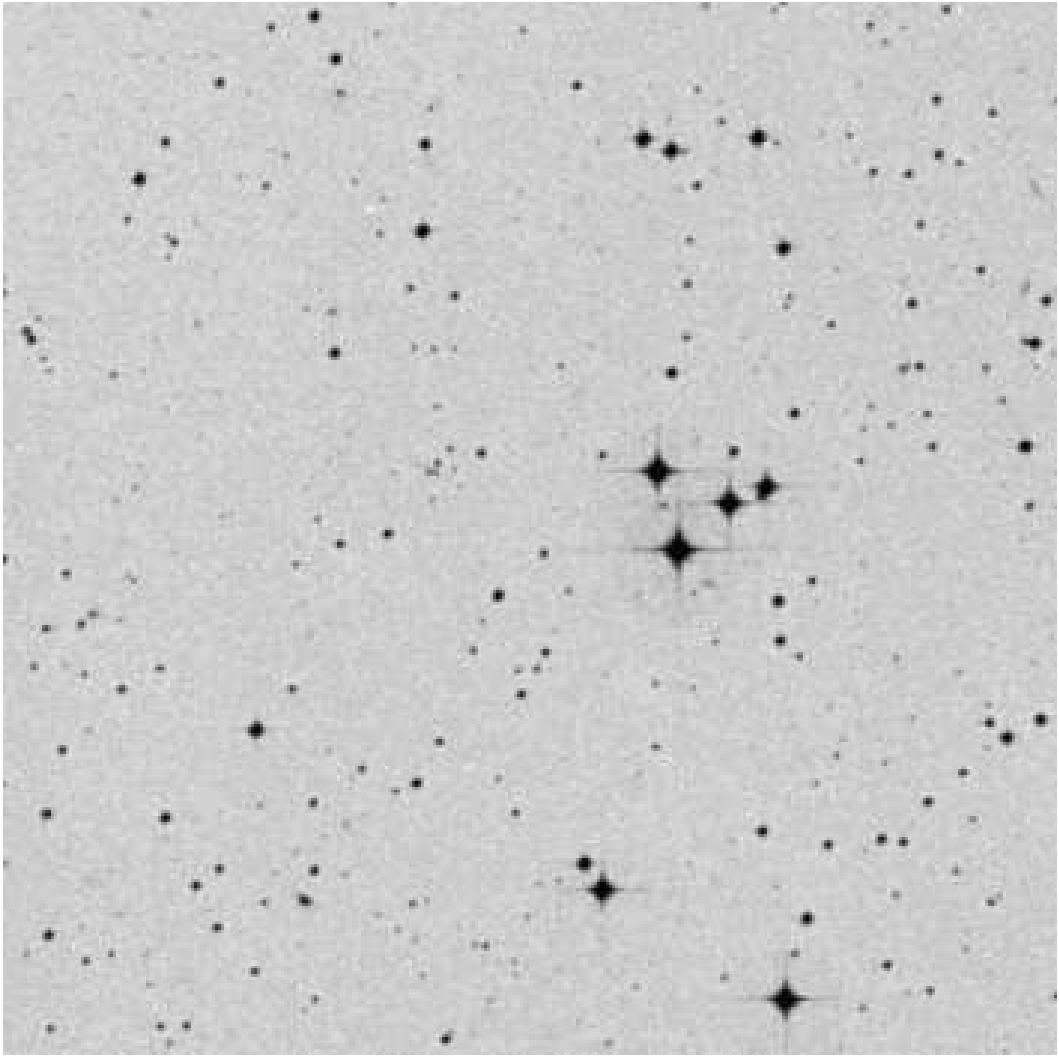,height=14cm,width=14cm}}
 \caption{A DSS $10 \times 10$ arcmin$^{2}$ image in the
 region of the open cluster NGC~6994. North is on the top,
 east on the left. The two brighter stars in the asterism
 are GSC~05778--0082 and HD~358033, respectively.}
\end{figure*}

In this paper we present $B$, $V$ and $I$ photometry for NGC~6994,
an object which appears in the Lyng{\aa} (1987) open clusters catalogue
and that it is classified as open cluster also by SIMBAD. \\
The aim is to provide some photometric data for this cluster, 
complementing the very poor informations we have, which basically consists 
of the cluster identification and diameter, which according to Lyng{\aa}
(1987) is about $1.0$ arcmin.
Its coordinates for the 2000.0 equinox are:
$\alpha~=~20^{h}~59^{m}~00^{s}$,
$\delta~=~-12^{o}~38{\prime}$,
$l~=~35^{o}.73$,
$b~=~-33^{o}.95$.\\
NGC~6994 is expected to lie close to HD~358033, GSC~05778--0082,
and M~73. This latter might be a binary or multiple 
system according to SIMBAD,
but more probably represents all the asterism. \\
From the acquired data, it turns out that NGC~6994 is not a cluster,
but just a random enhancement of the four mentioned bright stars.\\
Section~2 is dedicated to a brief description of the data acquisition
and reduction; Section~3 deals with the cluster structure and CMD.
Finally Section~4 summarizes the results.

\begin{table*}
\tabcolsep 0.08truecm
\caption{Basic parameters of the two brighter stars in NGC~6994 region.
Magnitudes, proper motions and trigonometric parallax are
from  the Tycho catalogue. }
\begin{tabular}{cccccccccc} \hline
\multicolumn{1}{c}{Name} &
\multicolumn{1}{c}{$\alpha$(2000.0)} &
\multicolumn{1}{c}{$\delta$(2000.0)} &
\multicolumn{1}{c}{$B$}&
\multicolumn{1}{c}{$V$}&
\multicolumn{1}{c}{$\mu_{\alpha}$} &
\multicolumn{1}{c}{$\mu_{\delta}$} &
\multicolumn{1}{c}{Spectral type}  &
\multicolumn{1}{c}{$\pi$} &
\multicolumn{1}{c}{dist} \\
 & hh:mm:ss &($^{o}$)($\prime$)($\prime\prime$)& & &$\prime\prime/yr$& $\prime\prime/yr$& & mas & pc\\
\\
HD~358033       &  20:58:57.9&-12:37:45.9& 11.915 & 11.170& 0.0190&-0.006&F5& 7.40&135\\
GSC~05778-00802 &  20:58:56.7&-12:38:30.1& 11.688 & 10.427& 0.0075&-0.011&  &23.80&42\\
\hline
\end{tabular}
\end{table*}

\section[]{Data Acquisition and Reduction }
Observations were conducted at La Silla on 1999 October 8, using the Tektronix
2024$\times$ 2024 pixel CCD $\#$ 36 mounted in the red EMMI
arm  of the 3.6-m ESO NTT
telescope. 
The reader is referred to Maris et al (2000) for any detail about
data reduction and acquisition.\\
Magnitudes and colors of all the observed stars are available upon request, 
together with the frame 
coordinates ($X$ and $Y$) and the instrumental ALLSTAR rms errors $\sigma$.

\section[]{The cluster}

\subsection{Preliminary considerations}
NGC~6994 is a poorly known object. It is classified as a Trumpler (1930)
{\it IV~1~p} cluster, say a poorly populated (4 stars?) and  compact system. 
This classification however is doubtful.
Many observers detected this stars concentration, which was baptized also as
M~73, Cr~426, C~2056-128, and OCL~89 (Ruprecht 1966).
Its angular diameter
is estimated to range between 1 to 2.8 arcmin, which means that it should be
a rather compact cluster (Collinder (1931) suggested it is a globular).\\ 
It seems that all these suggestions converge to the conclusion that we have 
to consider NGC~6994 as a group of four bright stars shown in Fig.~1.

\subsection{The structure}
NGC~6994 is a cluster located at relatively high galactic latitude
($b=-33.95$).
It appears as a concentration of four bright stars in the DSS
image presented in Fig~1, although the barycenter of these stars is offset
with respect to the commonly reported
cluster center, which corresponds to the center of the image. 
These four stars are rather close, justifying
the small diameter reported by Lyng{\aa} (1987), who presumably referred
to these stars when describing NGC~6994 properties. 
In details, 
the two brighter stars are HD~358033 and GSC~05778--0082, and their parameters
are listed in Table~1, where magnitudes ($B$ and $V$), distance
and proper motions are from Tycho catalogue (H$\o$g et al 1998).
Typical errors affecting these magnitudes are 0.3-0.4 mag.
These two stars are well inside the solar neighborhood, and lie
135 and 42 pc from the Sun, respectively.
M~73 according to Simbad is a double or multiple star, but most probably represents
the whole asterism, and it could be considered as a duplicate name for
NGC~6994.
Moreover we noticed that the faintest star in the asterism might be a 
visual binary, or simply the  two stars are projected in the
same sky direction. It is however rather unlikely to imagine that this
binary system actually represents M~73. 
Apart from these four stars,
no many other stars are visible close to the cluster center
which could  justify the classification of this object as 
an open cluster. Indeed the surrounding field (see Fig.~1)
appears very smoothly populated.
Looking at the cluster structure it is reasonable to suggest
that these four stars are responsible for the by eye
identification of this aggregate as an open cluster. 

\begin{figure*}
\centerline{\psfig{file=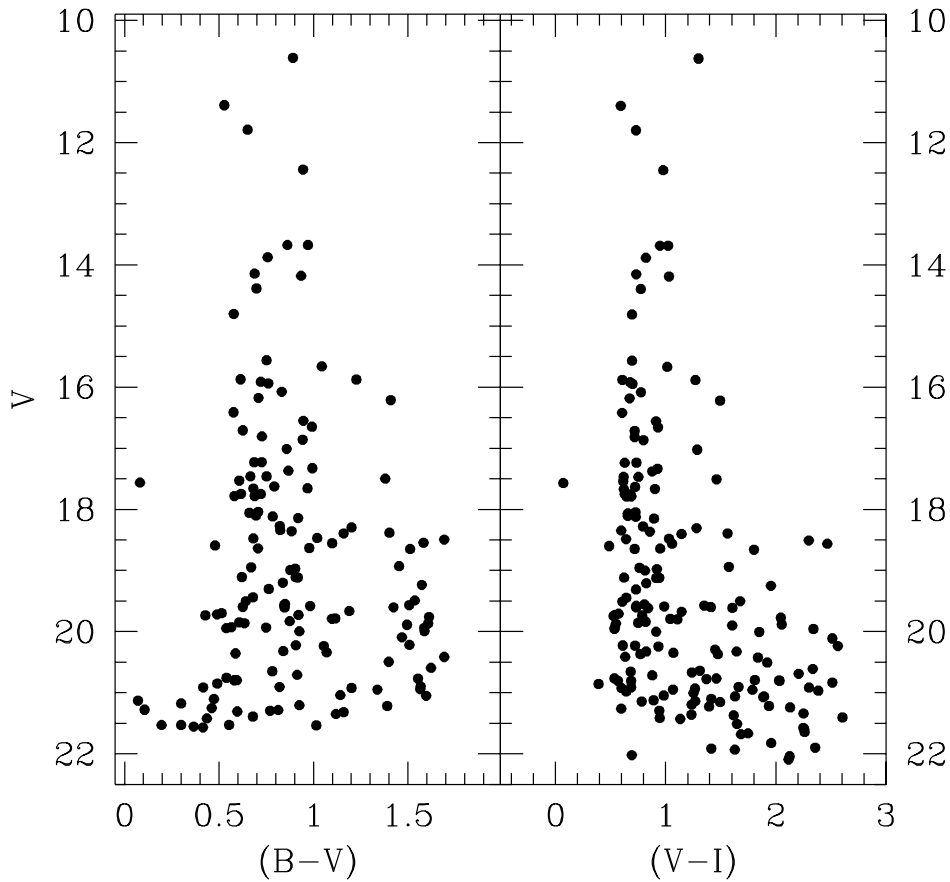,height=14cm,width=14cm}}
\caption{ The CMD of the measured stars in the region
of NGC~6994 in the V vs (V-I) (right panel) and V vs (B-V) (left panel) planes.}
\end{figure*}

\subsection{The CMD}
The measured stars in the plane $V$ versus  $(B-V)$ and
$V$ versus $(V-I)$ are shown in Fig.~2 (left and right panel, respectively),
and represent a region somewhat smaller ($9 \times 9$ arcmin$^{2}$)
than that shown in Fig.~1.
They define a broad vertical sequence which gets wider at increasing magnitude.
The distribution of stars does not exhibit any distinguishing feature, but
resembles a typical field stars CMD.
The straightforward conclusion is that NGC~6994 is not an open cluster,
but the stars in this region define a rather smooth field population,
with some voids and less rich in stars than the typical galactic disk fields.
We must stress that we are looking at a relatively high 
galactic latitude, where the thickness of the disk is rather small.

\subsection{Is NGC~6994 an open cluster remnant?}
de la Fuente Marcos (1998) studied the dynamical evolution of open star clusters,
suggesting that the final stage of their evolution consists of a handful of stars
which emerges from the general galactic field. The life-time, number of remaining stars
and dimension of the OCR depends on the initial cluster size and on the distance to the 
galactic center.
As for NGC~6994, we find that there are 11 stars which 
significantly emerge from the field (see Fig.~2). 
This would mean that NGC~6994 was a rich open 
cluster with an initial population of about 700 stars, and an age of almost a billion
yrs, or a younger (half a billion yrs) initially less rich (250 stars) 
open cluster if the binary population was significant ($30\%$).
Since NGC~6994 does not have bright stars
(the two brightest ones are probably dwarfs), the most plausible conclusion
would be that it was initially a rich populated cluster,  with an age of 1 Gyr
or more.
There are however not enough arguments leading to this scenario. 
In fact the lack of any feature in the CMD
is a strong argument against the classification
of this object as an open cluster. Indeed moving groups and OCR 
actually maintain in time 
some structures resembling a cluster CMD (see de la Fuente Marcos,
1998, Fig.~3). \\
The most reasonable conclusion is that we are looking at the general galactic field,
with stars at any distance from the Sun.

\section{Conclusions}
We reported on photometry in $B$, $V$ and $I$ passbands for NGC~6994, an object
previously classified as open cluster.\\
The analysis of the observed cluster field, and 
the distribution of the stars in the CMD seems 
to favor the suggestion that this object is not really an open cluster,
but simply an enhancement of four bright stars in the general galactic field.\\

\noindent
The conclusion of this work can be summarized as follows:

\begin{description}
\item[$\bullet$] NGC~6994 is an asterism of four stars and not an open cluster;
\item[$\bullet$] NGC~6994 is unlikely to be an OCR;
\item[$\bullet$] M~73 should not be considered  a binary system; 
instead  it can be used as a duplicate name for NGC~6994;
\end{description}

\noindent
Carraro \& Patat (1995) found another object classified as open cluster
- Ruprecht~46 - which was proved to be just a random
fluctuation in the field of the galactic disk.
We wonder whether
the possibility  exists that some other unstudied
or poorly studied open clusters might not be real open clusters, but
OCRs or just random star concentration fluctuations in the galactic disk.

\section*{Acknowledgments}
I thank Michele Maris and Gabriele Cremonese for the use
of observing time in common.
This work made use of SIMBAD and
has been financed by italian MURST and ASI.

\label{lastpage}

\end{document}